\documentclass[seceq]{ptptex}

\usepackage{graphicx}

\newlength{\lslash}
\def\slash#1{\settowidth{\lslash}{$#1$}\makebox[\lslash]{\makebox[0mm]{$/$}\makebox[0mm]{$#1$}}}

\title{
Self consistency in hadron physics }

\author{
M.F.M. \textsc{Lutz}$^{1,}$\footnote{ e-mail address:
m.lutz@gsi.de}, C.L. \textsc{Korpa}$^{2}$ and A.
\textsc{Semke}$^{1}$
 }

\inst{
$^1$ Gesellschaft f\"ur Schwerionenforschung (GSI),\\
Planck Str. 1, 64291 Darmstadt, Germany\\
$^{2}$ Department of Theoretical Physics, University of Pecs,\\
Ifjusag u. 6, 7624 Pecs, Hungary \\}

\abst{
In this talk we discuss at hand of two examples the crucial role
played by self consistency in hadron physics. The first example
concerns  the quark-mass dependence of the baryon octet and
decuplet masses.  It is shown that within a self consistent
one-loop approach based on the chiral Lagrangian the 'mysterious'
quark-mass dependence of the $\Xi$ mass predicted by the MILC
collaboration may be recovered in terms of a discontinuous chiral
extrapolation. This is a consequence of self consistency imposed
on the partial summation, i.e. the masses used in the loop
functions are identical to those obtained from the baryon self
energies. In the second example we discuss  recent studies on the
properties of D mesons in cold nuclear matter as they are
predicted by coupled-channel dynamics. Here a self consistent
many-body approach reveals the close interlink of the properties
of  D meson and open-charm baryon resonances in nuclear matter.
The relevance of exotic baryon resonances for the spectral
distortion of the $D_s^\pm$ in nuclear matter is pointed out. }

\begin{document}

\maketitle

\section{Introduction}

In this talk we discuss two seemingly unrelated topics. First the
quark-mass dependence of baryon masses and second the properties
of D mesons in cold nuclear matter. The two topics  exemplify the
crucial importance of self consistency in hadron physics. As it
will become clear during the talk both topic have exotic aspects
justifying its presentation in a workshop about 'exotics'.

The present-day interpretation of QCD lattice simulations requires
a profound understanding of the dependence of observable
quantities on the light quark masses. A powerful tool to derive
such dependencies is the chiral Lagrangian, an effective field
theory based on symmetry properties of QCD. The application of
strict chiral perturbation theory to the SU(3) flavor sector of
QCD is plagued by poor convergence properties for processes
involving baryons
\cite{Jenkins:1992,Ellis:Torikoshi:1999,Lehnhart:Gegelia:Scherer:2005}.
Thus it is important to establish partial summation schemes that
enjoy improved convergence properties and that are better suited
for chiral extrapolations of lattice simulations. We review the
results of \cite{Semke:Lutz:2005,Semke:Lutz:2006} where  recent
lattice QCD simulation of the MILC collaboration
\cite{MILC:2001,MILC:2004}, that use dynamical u-,d- and s-quarks
in the staggered approximation are interpreted. Adjusting the
values of the $Q^2$ counter terms to the physical masses we
suggest the possibility of a discontinuous dependence of the
baryon masses on the pion mass. This is a consequence of self
consistency imposed on the partial summation approach, i.e. the
masses used in the loop function are identical to those obtained
from the baryon self energy. The latter is a crucial requirement
since the loop functions depend sensitively on the precise values
of the baryon masses. Our results may explain the mysterious
quark-mass dependence for the $\Xi$ mass observed by the MILC
collaboration \cite{MILC:2001,MILC:2004}.

Except for the recent works
\cite{Tolos:Schaffner:Mishra:2004,Lutz:Korpa:2005,Mizutani:Ramos:2006}
a mean field ansatz was assumed to predict the properties of D
mesons in nuclear matter
\cite{Tsushima:1999,Sibirtsev:1999,Hayashigaki:2000,Tolos:Schaffner:Mishra:2004}.
This development resembles to a large extent the first attempts to
predict the properties of kaons and antikaons in nuclear matter
(see e.g. \cite{Kaplan:Nelson,njl-lutz:a,njl-lutz:b}), which also
assumed a mean-field type behavior for the mass shifts in nuclear
matter at first. However, by now it is well established that for
the antikaon such an ansatz is not valid due to complicated
many-body dynamics induced by the presence of resonance-hole
states \cite{ml-sp,Lutz:Korpa:2002}. A self consistent approach is
required that treats the spectral distortions of the hyperon
resonances and the antikaon in a consistent manner. As illustrated
in \cite{Lutz:Korpa:2005} it is crucial to study also the
properties of D mesons in a more microscopic manner. The key
ingredient of a realistic description are the D-meson nucleon
scattering amplitudes, that determine the D meson self energy at
least for dilute nuclear matter \cite{dover,njl-lutz:b}. In
contrast to the $D_-$-nucleon, the $D_+$-nucleon  system may
couple to open-charm baryon resonances. This poses a particular
challenge since the spectrum of the open-charm baryon resonances
is studied experimentally and theoretically poorly so far. As
suggested first in \cite{Lutz:Kolomeitsev:2004} the well
established  $\Lambda_c(2594)$ s-wave resonance resonance may be
dynamically generated by coupled-channel interactions
\cite{Tolos:Schaffner:Mishra:2004}. A detailed study
\cite{Hofmann:Lutz:05} that took into account charm-exchange
reactions systematically revealed that it couples strongly to the
$D_-$-nucleon channel. Thus the $\Lambda_c(2594)$  should play an
important role in the description of the nuclear $D_-$ dynamics.
Besides the narrow $\Lambda_c(2594)$, which couples dominantly to
the $D N, D_s \Lambda$ channels, the work \cite{Hofmann:Lutz:05}
recovers a broad s-wave state, so far unobserved, that is
interpreted as a chiral excitation of the open-charm sextet ground
states \cite{Lutz:Kolomeitsev:2004}. The latter couples strongly
to the $\pi \Sigma_c$ channels but very weakly to the $D N$
channel. It is the analogue of the $\Lambda(1405)$, which is a
chiral excitation of the baryon octet ground states
\cite{Wyld,Dalitz,Granada,Copenhagen}. A further striking
prediction of the work \cite{Hofmann:Lutz:05} is a narrow isospin
one resonance of mass 2.62 GeV which couples dominantly to the $D
N, D_s \Sigma$ channels. This state is so far unobserved, but, so
it existed, would affect the properties of $D_+$ mesons in nuclear
matter significantly \cite{Lutz:Korpa:2005}.

\section{Chiral extrapolation of baryon masses}

We collect the terms of the chiral Lagrangian that determine the
leading orders of baryon octet and decuplet self energies
\cite{Krause:1990}. Up to chiral order $Q^2$ the baryon
propagators follow from
\begin{eqnarray}
{\mathcal L} &=&  \mathrm{tr} \,\Big( \bar{B}\,\big[i\,
\slash{\partial}\,- \stackrel{\circ}{M}_{[8]}\big]\,B \Big)
\nonumber \\
&-& \mathrm{tr}\,\Big(\bar{\Delta}_\mu \cdot
\Big(\big[i\,\slash{\partial}\,
-\stackrel{\circ}{M}_{[10]}\big]\,g^{\mu\nu} -i\,(\gamma^\mu
\partial^\nu + \gamma^\nu \partial^\mu) +
\gamma^\mu\,\big[i\,\slash{\partial} +
\stackrel{\circ}{M}_{[10]}\big]\,\gamma^\nu \Big)\,\Delta_\nu\Big)
\nonumber \\
& -& 2\,d_0\, \mathrm{tr}\Big(\bar{\Delta}_\mu \cdot \Delta^\mu
\Big)\, \mathrm{tr}\Big(\chi_0\Big) -2\,d_D\, \mathrm{tr}\Big(
(\bar{\Delta}_\mu \cdot \Delta^\mu)\, \chi_0\Big)
\nonumber \\
&+&2\,b_0 \,\mathrm{tr}\Big(\bar{B}\,B\Big)\,
\mathrm{tr}\Big(\chi_0\Big) +
2\,b_F\,\mathrm{tr}\Big(\bar{B}\,[\chi_0,B]\Big) +
2\,b_D\,\mathrm{tr}\Big(\bar{B}\,\{\chi_0,B\}\Big) \,,
\nonumber\\
\nonumber\\
&& \chi_0 = \left( \begin{array}{ccc}
m_\pi^2 & 0 & 0 \\
0 & m_\pi^2 & 0 \\
0 & 0 & 2\,m_K^2-m_\pi^2
\end{array}\right)\,,
\label{chiral-L}
\end{eqnarray}
where we use the notations of
\cite{Semke:Lutz:2005,Semke:Lutz:2006} for the baryon octet and
decuplet fields $B$ and $\Delta$ respectively.  The evaluation of
the baryon self energies to order $Q^3$ probes the meson-baryon
vertices
\begin{eqnarray}
{\mathcal L} &=& \frac{F}{2f}\, \mathrm{tr} \,\Big( \bar{B}\,
\gamma_5 \gamma^\mu \,[\partial_\mu \Phi,\,B] \Big) +
\frac{D}{2f}\, \mathrm{Tr}\,\Big( \bar{B}\, \gamma_5 \gamma^\mu
\{\partial_\mu \Phi,B\} \Big)
\nonumber\\
&-& \frac{C}{2f}\, \mathrm{tr}\,\Big(\bar{\Delta}_\mu \cdot
(\partial_\nu \Phi)\, \big[g^{\mu\nu}-\frac{1}{2}\,Z\,\gamma^\mu
\,\gamma^\nu\big]\, B + \mathrm{h.c.} \Big)
\nonumber\\
&-& \frac{H}{2f} \mathrm{tr} \,\Big(\big[ \bar{\Delta}^\mu \cdot
\gamma_5\,\gamma_\nu\, \Delta_\mu \big]\,(\partial^\nu
\Phi)\,\Big)\,, \label{chiral-FD}
\end{eqnarray}
where we apply the notations of \cite{Lutz:Kolomeitsev:2002}. We
use $f = 92.4$ MeV  in this work. The values of the coupling
constants $F,D,C$ and $H$ may be correlated by a large-$N_c$
operator analysis \cite{Dashen,Jenkins,Jenkins:Manohar}. At
leading order the coupling constants can be expressed in terms of
$F$ and $D$ only. We employ the values for $F$ and $D$ as
suggested in \cite{Okun,Lutz:Kolomeitsev:2002}. All together we
use
\begin{eqnarray}
&& F = 0.45 \,, \qquad D= 0.80 \,, \qquad
 H= 9\,F-3\,D \,,\qquad C=2\,D \,.
\label{large-Nc}
\end{eqnarray}
We take the parameter $Z=0.72$ from a detailed coupled-channel
study of meson-baryon scattering that was based on the chiral
SU(3) Lagrangian \cite{Lutz:Kolomeitsev:2002}. The latter
parameter is an observable quantity within the chiral SU(3)
approach: it contributes at order $Q^2$ to the meson-baryon
scattering amplitudes and cannot be absorbed into the available
$Q^2$ counter terms \cite{Lutz:Kolomeitsev:2002}.

It is straight forward to evaluate the one-loop contributions to
the baryon self energy as implied by (\ref{chiral-FD}). Explicit
and complete expressions are provided for the first time in
\cite{Semke:Lutz:2005,Semke:Lutz:2006}. Since we do not advocate a
strict chiral expansion, rather a partial summation, the
renormalization needs to be discussed briefly. As pointed out in
\cite{Semke:Lutz:2005}, within dimensional regularization there
persists an ambiguity  on how to implement the chiral counting
rules. This leads to the presence of an infrared renormalization
scale that can be used to optimize the speed of convergence.
Performing a strict chiral expansion the physical parameters are
independent on the infrared scale, as they should be. However, the
size of the $Q^2$ counter terms depend linearly on this scale. In
turn the apparent convergence properties reflect the choice of
that scale. Only for reasonable values of the infrared scale,
$\mu_{IR}$,  the counter terms have a natural size.

Once a partial summation, as implied by a self consistent
evaluation of the loop function, is performed  an infinite number
of counter terms would be needed to arrive at results that are
manifestly independent on the renormalization scales. Dialing a
value of the infrared scale amounts to a particular choice
thereof. Any residual dependence of the baryon masses on the
infrared scale may be used to estimate the uncertainty of the
given truncation.

\begin{table}[t]
\begin{center}
\begin{tabular}{|c|c|c|c|}\hline
 & $\mu_{IR}=350$\, MeV &  $ \mu_{IR}=450$\, MeV &  $ \mu_{IR}=550$\,
MeV \\ \hline \hline
$b_0\; \mathrm{[GeV^{-1}]}$ & $-0.89$ & $-0.63$ & $-0.38$\\
$b_D\; \mathrm{[GeV^{-1}]}$ & $+0.29$ & $+0.19$ & $+0.10$\\
$b_F\; \mathrm{[GeV^{-1}]}$ & $-0.34$ & $-0.25$ & $-0.15$\\
$d_0\; \mathrm{[GeV^{-1}]}$ & $-0.22$ & $-0.15$ & $-0.08$\\
$d_D\; \mathrm{[GeV^{-1}]}$ & $-0.35$ & $-0.30$ & $-0.24$\\
\hline \hline $M_N$ [MeV] &
$\phantom{1}750+\phantom{1}310-\phantom{1}121  $& $
\phantom{1}813+\phantom{1}232-\phantom{1}105 $ &
$ \phantom{1}875+\phantom{1}153-\phantom{11}89 $\\
 & $=\phantom{1}939  $& $ =\phantom{1}939 $ & $ =\phantom{1}939 $\\
$M_\Lambda$ [MeV] & $\phantom{1}750+\phantom{1}536-\phantom{1}150
$&$ \phantom{1}813+\phantom{1}398-\phantom{111}79 $
&$ \phantom{1}875+\phantom{1}260-\phantom{111}9 $\\
 & $=1136  $&$ =1131 $ &$ =1126 $\\
$M_\Sigma$ [MeV] & $\phantom{1}750+\phantom{1}882-\phantom{1}425 $
& $\phantom{1}813+\phantom{1}630-\phantom{1}239 $
& $\phantom{1}875+\phantom{1}379-\phantom{11}54 $\\
 & $=1207 $ & $=1203 $ & $=1200 $\\
$M_\Xi$ [MeV] & $\phantom{1}750+\phantom{1}934-\phantom{1}361 $&
$\phantom{1}813+\phantom{1}680-\phantom{11}171 $
& $\phantom{1}875+\phantom{1}427+\phantom{11}19 $\\
& $=1323 $& $=1321 $ & $=1320 $\\
\hline \hline $M_\Delta$ [MeV] &
$1082+\phantom{1}241-\phantom{11}91  $&
$1108+\phantom{1}164-\phantom{11}39$
& $1133 +\phantom{11}87+\phantom{11}12$\\
 & $=1232  $& $ =1232 $ & $ =1232 $\\
$M_\Sigma$ [MeV] & $1082+\phantom{1}347-\phantom{11}49  $&$
1108+\phantom{1}253+\phantom{11}17 $
&$1133+\phantom{1}160+\phantom{11}83 $\\
 & $=1380  $&$ =1378 $ &$ =1376 $\\
$M_\Xi$ [MeV] & $1082+\phantom{1}453-\phantom{111}4 $ &
$1108+\phantom{1}343+\phantom{11}78 $
& $1133+\phantom{1}233+\phantom{1}160 $\\
 & $=1530 $ & $=1528 $ & $=1526 $\\
$M_\Omega$ [MeV] & $1082+\phantom{1}558+\phantom{11}34 $&
$1108+\phantom{1}432+\phantom{1}134$
& $1133+\phantom{1}305+\phantom{1}235$\\
 & $=1674 $& $=1674 $ & $=1674 $\\
\hline \hline
$\sigma_{\pi N}$ [MeV] & 52.4 & 53.9 & 55.7\\
\hline
$\sigma_{K^- p}$ [MeV] & 384.0 & 380.1 & 386.3\\
\hline $\sigma_{K^- n}$ [MeV] & 359.4 & 354.8 & 361.5\\ \hline
\end{tabular}
\caption{The parameters are fitted to reproduce the baryon masses
at physical pion masses as well as the SU(3) limit values $M_{[8]}
\simeq 1575$ MeV and $M_{[10]}\simeq 1710$ MeV at $m_{\pi}\simeq
690$ MeV. We use $\mu_{UV}=800$ MeV. The masses are decomposed
into their chiral moments. } \label{tab:parameter}
\end{center}
\end{table}

\begin{figure}[tbh]
\centering
\includegraphics[width=8.0cm]{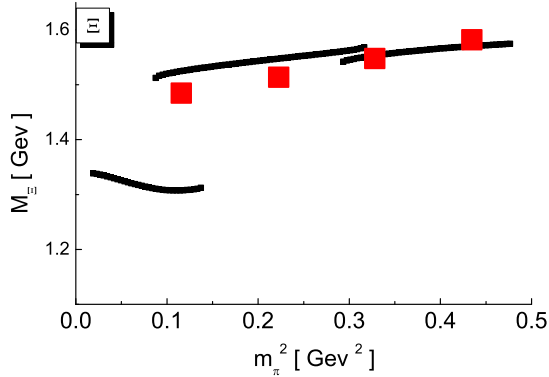}
\caption{The pion mass dependence of baryon octet and decuplet
masses predicted by the chiral loop expansion taking the
parameters of Tab. \ref{tab:parameter}. The  lines represent the
masses for the infrared scale put at $\mu_{IR}=450$ MeV. The solid
squares are the simulation points of the MILC collaboration.
}\label{fig:shape}
\end{figure}

For given values of the infrared and ultraviolet renormalization
scales the parameters $b_D, b_F$ and $d_D$ are fitted to the mass
differences of the octet states and decuplet states. The absolute
mass scale of the octet and decuplet states can be reproduced by
appropriate values of the bare baryon masses. This procedure
leaves undetermined the two parameters $b_0$ and $d_0$. Good
representations of the physical baryon masses can be obtained for
a wide range of the latter. The parameter $b_0$ may be used to
dial a given pion-nucleon sigma term at physical pion masses.
Similarly the unknown parameter $d_0$ may be determined to
reproduce a given pion-delta sigma term. The latter two parameters
are adjusted as to reproduce the baryon octet and decuplet masses
in the SU(3) limit at $m_{\pi}\simeq 690$ MeV. The MILC
simulations suggest the values $M_{[8]} \simeq 1575$ MeV and
$M_{[10]}\simeq 1710$ MeV. This procedure is biased to the extent
that it assumes that the self consistent one-loop results will be
applicable at such high quarks masses. On the other hand as long
as there are no continuum limit results of the MILC collaboration
available this is an economical way to minimize the influence of
lattice size effects. The latter are expected to be smaller at
large quark masses. The procedure may be justified in retrospect
if it turns out that the extrapolation recovers the behavior
predicted by the lattice simulation.

The parameters used are collected in Tab. \ref{tab:parameter}
together with the implied masses of the baryon octet and decuplet
states. A fair representation of the physical baryon masses is
obtained.  As discussed in detail in \cite{Semke:Lutz:2005} the
parameters $b_{0,D,F}$ and $d_{0,D}$ show a strong dependence on
the infrared scale $\mu_{IR}$. In contrast, the physical baryon
masses suffer from a weak dependence only. For natural values of
the infrared scale the chiral expansion appears well converging as
indicated by the decomposition of the baryon masses into their
moments. Taking the residual scale dependence of the pion-nucleon
sigma term as a naive error estimate we obtain $\sigma_{\pi N} =
54 \pm 2$ MeV.

We turn to the pion-mass dependence of the baryon octet and
decuplet masses. It should be emphasized that the baryon masses
are a solution of a set of coupled and non-linear equations in the
present scheme. This is a consequence of self consistency imposed
on the partial summation approach.  As a consequence of the
non-linearity  for a given parameter set there is neither a
guarantee for a unique solution to exist, nor that solutions found
are continuous in the quark masses. Indeed, as illustrated by Fig.
1 the pion-mass dependence predicted by the chiral loop expansion
is quite non-trivial exhibiting various discontinuities. It is
striking to see that we reproduce the 'mysterious' pion-mass
dependence of the $\Xi$ mass, i.e. the quite flat behavior which
does not seem to smoothly approach the physical mass. Given the
present uncertainties from finite lattice spacing, the staggered
approximation and the theoretical uncertainties implied by higher
order contributions, we would argue that we arrive at a fair
representation of the lattice simulation points for all baryons
(see \cite{Semke:Lutz:2006}) with some reservation concerning the
nucleon. Incorporating the many $Q^4$ counter terms offered by the
chiral Lagrangian it is reasonable to expect that the latter will
further improve the picture. However, as long as there is no
detailed analysis available that performs the continuum limit
there is not much point considering the $Q^4$ counter terms.

\section{D mesons in nuclear matter}

After a demonstration of the crucial importance of self
consistency for the chiral extrapolation of the baryon octet and
decuplet masses we turn to the propagation of D mesons in a cold
nuclear environment. We discuss the properties of the $D^\pm$ and
$D^\pm_s$ mesons as derived from a self consistent many body
approach based on coupled-channel dynamics \cite{Lutz:Korpa:2005}.
The dominant interaction was modelled by the exchange of light
vector mesons in the t-channel \cite{Hofmann:Lutz:05}. All
relevant coupling constants were obtained from chiral and
large-$N_c$ properties of QCD. Less relevant three-point vertices
related to the t-channel forces induced by the exchange of charmed
vector mesons were estimated by a flavor SU(4) ansatz
\cite{Hofmann:Lutz:05}. The resulting s-wave $D_-N \to D_- N$
scattering amplitudes are dominated by the dynamically generated
$\Lambda_c(2594)$ and $\Sigma_c(2620)$ resonances. The s-wave
$D^\pm_s N \to D^\pm_s N$ scattering amplitudes are characterized
by so far unobserved exotic resonances at 2.89 GeV and 2.78 GeV.
Only the $D_- N \to D_- N$ scattering process is not influenced by
the presence of a resonance. In this case the amplitude is
characterized to a large extent by the scattering length
\begin{eqnarray}
&& a^{(I=0)}_{D^-N} \simeq-0.16\, {\rm fm} \,, \qquad  \quad
a^{(I=1)}_{D^-N} \simeq -0.26\, {\rm fm} \,. \label{length}
\end{eqnarray}

\begin{figure}[b]
\centering
\includegraphics[width=10.0cm]{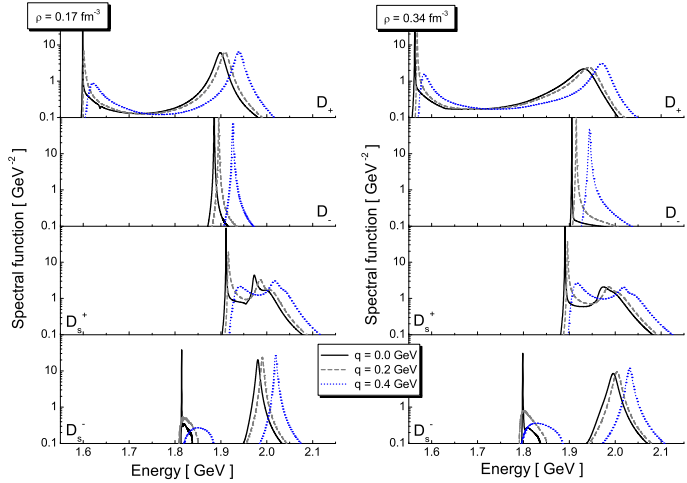}
\caption{Spectral distributions of the $D^\pm$ and $D_s^\pm$
mesons. Results are shown for the meson momenta 0, 200 and 400 MeV
and nuclear densities $0.17$ fm$^{-3}$ and $0.34$ fm$^{-3}$. The
self consistent many-body approach of \cite{Lutz:Korpa:2002} was
applied. }\label{fig:spectral:function}
\end{figure}

\begin{figure}[t]
\centering
\includegraphics[width=8.0cm]{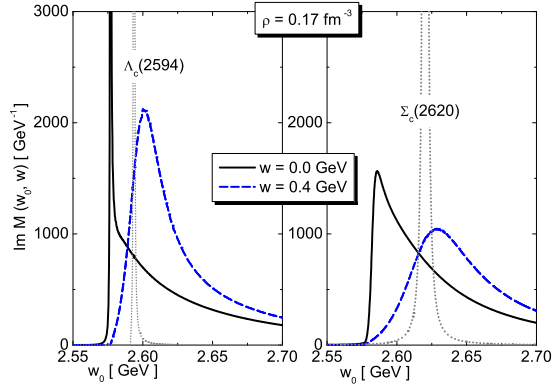}
\caption{Spectral distributions of the $D^\pm$ and $D_s^\pm$
mesons. Imaginary part of the isospin zero (l.h.p.) and isospin
one (r.h.p) $D_+$-nucleon scattering amplitude at saturation
density as compared to the free-space case (dotted lines). The
amplitudes are shown for two values of the resonance
three-momentum  $w=0$ GeV and $w=0.4$ GeV.}\label{fig:resonances}
\end{figure}

The many-body computations are based on the self consistent and
covariant formalism established in \cite{ml-sp,Lutz:Korpa:2002}.
It is important to perform such computations in a self consistent
manner since the feedback of an altered meson spectral function on
the resonance structure is typically an important effect
\cite{ml-sp,Lutz:Korpa:2002,Korpa}. Numerical results for the
spectral distributions of the D mesons as derived in
\cite{Lutz:Korpa:2005} are shown in Fig.
\ref{fig:spectral:function}. Given the scattering lengths
(\ref{length}) the mass shift of a $D_-$ meson in nuclear matter
is fully determined by the low-density theorem
\cite{dover,njl-lutz:b}. The low-density mass shift of 17 MeV is
quite close to the self consistent result shown in Fig.
\ref{fig:spectral:function}. Self-consistency leads to a mass
shift of 18 MeV at saturation density and 38 MeV at twice
saturation density. The spectral function of the $D_+$ meson has a
two-mode structure, which is a consequence of important
resonance-hole contributions. At saturation density the main mode
is pushed up by about 32 MeV as compared to the free-space meson.
Our results for the $D_+$ meson differ from the previous study
\cite{Tolos:Schaffner:Mishra:2004} significantly. This is a
consequence of the quite different interaction used in the two
computations. The work \cite{Tolos:Schaffner:Mishra:2004}
overestimates the charm-exchange channels. In
\cite{Hofmann:Lutz:05} those channels are suppressed by a
kinematical factor $m^2_\rho/m^2_D \sim 0.2 $. In particular the
work \cite{Tolos:Schaffner:Mishra:2004} did not predict the
isospin one resonance $\Sigma_c(2620)$. The latter dominates the
resonance-hole component in the spectral distribution.

We turn to the mesons with non-zero strangeness. The study of
\cite{Hofmann:Lutz:05} predicts the existence of a coupled-channel
molecule in the ($D^-_s N,\, \bar D \Lambda,\, \bar  D \Sigma$)
system. The latter carries exotic quantum numbers that can not be
arranged by three quarks only. Exotic s-wave states with C=-1 were
discussed first by Gignoux, Silvestre-Brac and Richard
\cite{Gignoux:Silvestre-Brac:Richard} and later by Lipkin
\cite{Lipkin:87}. The binding of about 190 MeV for such a the
state as predicted in \cite{Hofmann:Lutz:05} awaits experimental
confirmation. If confirmed the s-wave $D^- _s N \to D^-_s  N$
scattering amplitude must show a prominent pole structure at
subthreshold energies. The latter leads to a well separated
two-mode structure of the $D^-_s$ spectral distribution. The main
mode is pushed up by less than 10 MeV at nuclear saturation
density.

The in-medium spectral distribution of the $D^+_s$ is most
striking. The additional mode expected from the possible existence
of the exotic state at 2.89 GeV merges with the main $D^+_s$ mode
into one broad structure, in particular at intermediate meson
momenta 400 MeV. The results are quite analogous to the spectral
distribution of the $K_-$ where the $\Lambda(1405)$ nucleon-hole
state gives rise to a broad distribution
\cite{ml-sp,Lutz:Korpa:2002}. Clearly this result is an immediate
consequence of the small binding energy of the exotic state.

In Fig. \ref{fig:resonances} the spectral functions of the
$\Lambda_c(2594)$ and $\Sigma_c(2620)$ resonances are shown at
saturation density as compared to their free-space distributions.
We observe small attractive shifts in their mass distributions and
significant broadening, at least once the resonances move relative
to the matter bulk. The shifts are the result of a subtle balance
of the repulsive Pauli blocking effect and attraction implied by
the coupling of the $D_+$ meson to resonance-hole modes.

\section{Conclusions}

The pion-mass dependence of the baryon octet and decuplet masses
was discussed. A novel approach was reviewed in which the latter
are a solution of a set of coupled and non-linear algebraic
equations. This is a direct consequence of self consistency
imposed on the partial summation, i.e. the masses used in the loop
functions are identical to those obtained from the baryon self
energies. As a striking consequence a discontinuous dependence of
the baryon masses on the pion mass arises. Typically the baryon
masses jump at pion masses as low as 300 MeV. Most spectacular is
the behavior of the $\Xi$ mass. At small pion masses it decreases
with increasing pion masses. At a critical pion  mass of about
300-400 MeV it jumps up to a value amazingly close to the
prediction of the MILC collaboration.

In  the second part of this talk we discussed recent results on
the properties of $D_\pm$ and $D^\pm_s$ mesons in cold nuclear
matter based on self consistent coupled-channel dynamics. It was
pointed out that the $D_+,D_s^\pm$ spectral distributions are
strongly distorted in a nuclear medium due to the presence of
resonances that couple strongly to the $D_+,D_s^\pm$ nucleon
channel. Since even the existence of most resonances that are
predicted by coupled-channel dynamics is not yet confirmed
experimentally such studies suffer from large uncertainties. This
asks for a dedicated experimental programm to unravel the spectrum
of baryons with charm content.


\bibliography{literature}
\bibliographystyle{h-physrev}

\end{document}